\documentclass[aps,reprint,prr,twocolumn,floatfix,amsmath,amssymb]{revtex4}
\usepackage[dvipdfm]{graphicx}
\usepackage{color}
\usepackage{dcolumn}
\begin{document}
\preprint{}
\title{
Hydrodynamic interactions in anomalous rheology of active suspensions
}
\author{Haruki Hayano and Akira Furukawa}
\thanks{furu@iis.u-tokyo.ac.jp}
\affiliation{Institute of Industrial Science, 
University of Tokyo, Meguro-ku, Tokyo 153-8505, Japan}
\date{\today}
\begin{abstract}
We explore a mechanism of the anomalous rheology of active suspensions by hydrodynamic simulations using model pusher swimmers.  
Our simulations demonstrate that hydrodynamic interactions under shear flow systematically orient swimmers along the extension direction, which is responsible for determining the global swimming states and the resulting significant viscosity reduction. 
The present results indicate the essential role of hydrodynamic interactions in the elementary processes controlling the rheological properties in active suspensions. Furthermore, such processes may be the substance of the previously proposed scenario for anomalous rheology based on the interplay between the rotational diffusivities and the external shear flow.
\end{abstract}

\maketitle
\section{Introduction}

Anomalous rheology observed in the broad class of active suspensions is one of the most typical phenomena highlighting distinctive differences from passive systems \cite{Hatwalne,Sokolov,Gachelin,Lopez,Liu,PNAS2020,Rafai,Review2,Marchetti,Review3}
. 
In particular, for rod-like extensile pusher micro-swimmers (such as {\it E. coli}), a significant viscosity reduction has been experimentally observed at lower shear rates and volume fractions \cite{Sokolov,Gachelin,Lopez,Liu,PNAS2020}, which frequently leads to a superfluid state with zero viscosity \cite{Lopez,PNAS2020}. A seminal study by Hatwalne {\it et al.} \cite{Hatwalne} predicted that if an orientational {\it order} along the extension axis of the applied flow is somehow realized, the active dipolar forces intensify the mean flow, reducing the resistive stress required to drive the external flow and thus the viscosity. Following that, many theoretical attempts have been made to predict or explain the anomalous rheology in active suspensions  (see papers \cite{Ishikawa_Pedray,Cates,Haines,Giomi,Saintillan1,Ryan,Moradi_Najafi,Nechtel_Khair,Takatori_Brady} and the references therein). 

In dilute suspensions of rod-like particles, the orientational distribution of particles under shear flow are known to be enhanced along the extension axis when (thermal or athermal) random rotational diffusion processes exist \cite{Hinch_Leal1,Hinch_Leal2}. By taking such fluctuation effects into account, the viscosity reduction was successfully modeled within the framework of continuum kinetic theory \cite{Haines,Saintillan1}. 

In dilute/semidilute active suspensions, hydrodynamic interactions (HIs) are expected to play a crucial role in couplings among constituents \cite{Review1,LaugaB}. In Ref. \cite{Ryan}, it is theoretically demonstrated that the long-range HIs induce marked density fluctuations that provide additional sources of the effective rotational noise, resulting in a decrease in the viscosity. Indeed, recent experiments \cite{PNAS2020} indicate a close link between collective many-body properties and anomalous rheology. Nevertheless, due to the highly nonlinear and nonequilibrium nature of HIs, our understanding of the extent to which interactions among swimmers are involved in the rheological properties is still lacking beyond the effective one-body theory. 

In this study, we investigate the mechanism of the anomalous viscosity reduction observed in active suspensions by revisiting the role of HIs. Our analysis, along with a phenomenological explanation, elucidates that swimming along the extension axis of the applied flow is hydrodynamically favorable, resulting in a significant reduction of the viscosity. Furthermore, we argue that in usual swimming bacteria, such as {\it E. coli}, the self-propulsive forces are strong enough that the induced HIs can compete or dominate other effects like thermal fluctuations even in dilute suspensions.

\section{Model swimmer system}

For the present purpose, we perform hydrodynamic simulations of model active suspensions composed of $N$ rod-like dumbbell swimmers with a prescribed force dipole. 
Our model swimmer, schematically shown in Fig. 1(a), is composed of body and flagellum parts. 
The body part is treated as a rigid body, while the flagellum part is regarded as a massless ``phantom'' particle simply following the body's motions. 
This treatment always keeps the relative position of these two parts unchanged.  
For the $\alpha$-th swimmer ($\alpha=1,\cdots, N$), it is assumed that the force $F_A{\hat {\mbox{\boldmath$n$}}}_{\alpha}$ acting on the (front) body is exerted by the (rear) flagellum  and that the flagellum also exerts the force $-F_A{\hat {\mbox{\boldmath$n$}}}_{\alpha}$ directly on the solvent fluid. 
Here, ${\hat {\mbox{\boldmath$n$}}}_{\alpha}$ is the direction of the $\alpha$-th swimmer, and these forces compose a dipolar force (please refer to Appendix A for details). 
The present particle-base model is essentially the same as those proposed in Refs. \cite{Graham1,Graham2} and used in Refs. \cite{Haines,Ryan,Gyrya,Decoene,Furukawa_Marenduzzo_Cates}. 
Continuum kinetic models of hydrodynamically interacting rod-like swimmers with prescribed stresses or forces were also developed \cite{Haines,Ryan,Saintillan-Shelley,Saintillan-Shelley2,Saintillan-Shelley3,Baskaran-Marchetti}. 
In Refs. \cite{Saintillan-Shelley,Saintillan-Shelley2,Saintillan-Shelley3}, 
it was demonstrated that nonlinear hydrodynamic effects can lead to larger-scale correlated motions with marked density fluctuations. 

\begin{figure}[bth] 
\includegraphics[width=8.7cm]{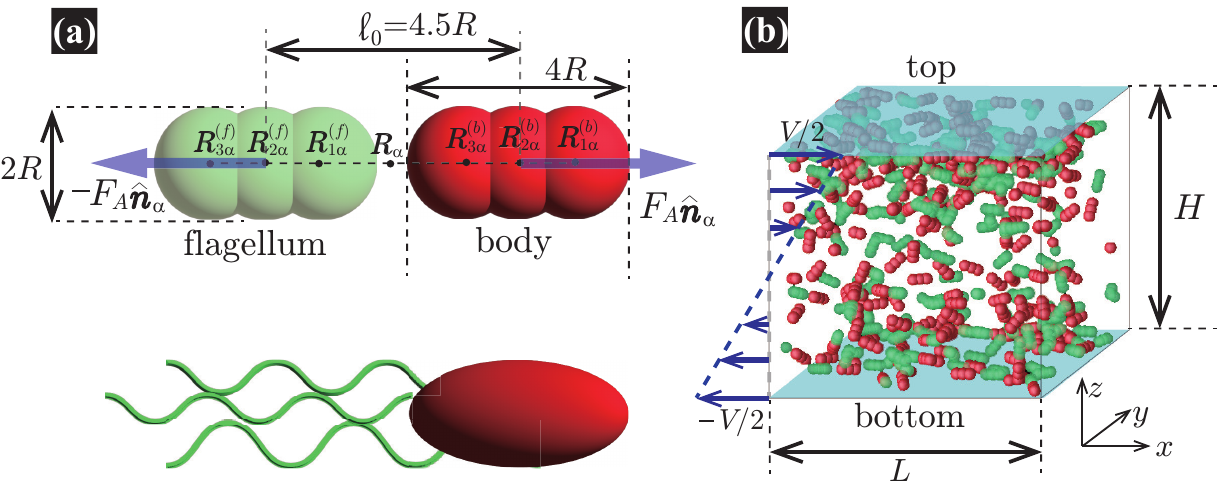}
\caption{(Color online) 
(a) Our model swimmer comprises ``body'' and ``flagellum'' parts with symmetric shapes. Each parts are constituted by a superposition of three spheres with radius $R$. 
We assume that the force $F_A{\hat {\mbox{\boldmath$n$}}}_\alpha$ is exerted on the body, while $-F_A{\hat {\mbox{\boldmath$n$}}}_\alpha$ is directly exerted on the solvent through the flagellum part, with ${\hat {\mbox{\boldmath$n$}}}_\alpha$ being the orientation of the $\alpha$-th swimmer. These forces constitute a force dipole with the magnitude $F_A \ell_0$. Here, $\ell_0$ is the characteristic swimmer's length and, for the present model, is given as the separation distance between the body and flagellum centers.  
(b) The periodic boundary conditions are imposed in the $x$- and $y$-directions with the linear dimension $L$.  
The shear flow is imposed by moving the top and bottom walls in the $x$-direction at constant velocities $V/2$ and $-V/2$, respectively. These two walls are separated in the $z$-direction by the distance $H$. 
}
\label{Fig1}
\end{figure}

As illustrated in Fig. \ref{Fig1}(a), the body and flagellum parts are assumed to have the same shape and are each described by a superposition of three spheres with a common radius $R$. 
The spheres composing the body are located at the positions  
${{\mbox{\boldmath$R$}}}_{i,\alpha}^{(b)}={{\mbox{\boldmath$R$}}}_{\alpha}^{(G)}+(2-i)R{\hat {\mbox{\boldmath$n$}}}_{\alpha}$ ($i=1,2,3$), where ${{\mbox{\boldmath$R$}}}_{\alpha}^{(G)}$ is the $\alpha$-th swimmer's center-of-mass position. 
Similarly, the spheres composing the flagellum part are located at ${{\mbox{\boldmath$R$}}}_{i,\alpha}^{(f)}={{\mbox{\boldmath$R$}}}_{\alpha}^{(CF)}+(2-i)R{\hat {\mbox{\boldmath$n$}}}_{\alpha}$ ($i=1,2,3$), where ${{\mbox{\boldmath$R$}}}_{\alpha}^{(CF)}={{\mbox{\boldmath$R$}}}_{\alpha}^{(G)}-4.5R{\hat {\mbox{\boldmath$n$}}}_{\alpha}={{\mbox{\boldmath$R$}}}_{\alpha}^{(G)}-\ell_0{\hat {\mbox{\boldmath$n$}}}_{\alpha}$ is the position of the center of the flagellum. 
The shape of the present model swimmer shows the head-tail symmetry, and the mid point is thus given by ${{\mbox{\boldmath$R$}}}_{\alpha}=({{\mbox{\boldmath$R$}}}_{\alpha}^{(G)}+{{\mbox{\boldmath$R$}}}_{\alpha}^{(CF)})/2$. 
Although arbitrary shapes of swimmers with an imposed head-tail asymmetry can be composed, we may obtain qualitatively the same results as long as these swimmers have rod-like forms with the prescribed force dipoles.

Periodic boundary conditions are imposed in the $x$- and $y$-directions with the linear dimension $L$, and the planner top and bottom walls are placed at $z=H/2$ and $-H/2$, respectively. The shear flow is imposed by moving the top and bottom walls in the $x$-direction at constant velocities $V/2$ and $-V/2$, respectively, whereby the mean shear rate is $\dot\gamma=V/H$. This situation is illustrated in Fig. 1(b). Hydrodynamic interactions among the swimmers are incorporated by adopting the smoothed profile method (SPM) \cite{SPM,SPM2,SPM3}, which is one of the mesoscopic simulation techniques \cite{LB1,LB2,FPD,FPD2,MPC,DPD} . 
In the SPM \cite{SPM,SPM2,SPM3}, the dynamics of rigid particles and a host fluid can be considered simultaneously with vastly reducing numerical costs by replacing particle-fluid boundaries with smoothed ones and by taking particle rigidity into account through the body force term in the Navier-Stokes equation. 
The details of the simulation methods are presented in Appendix A and B. 

\section{Results}
\subsection{Steady-state properties: weak alignment of the swimmers and the resultant viscosity reduction}

\begin{figure}[hbt] 
\includegraphics[width=8.7cm]{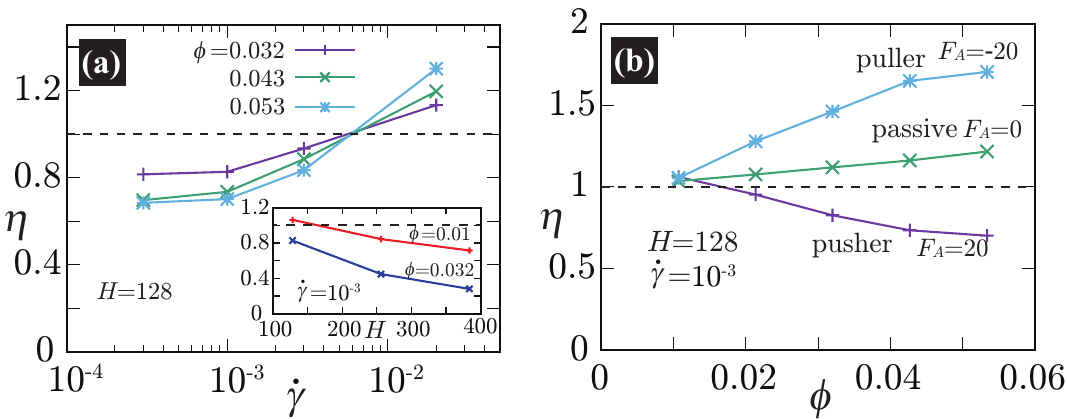}
\caption{(Color online) 
(a) The $\dot\gamma$-dependence of the viscosity $\eta$ for various $\phi$ at $H=128$.  
For smaller $\dot\gamma$ and larger $\phi$, $\eta$ becomes smaller. The inset shows $H$-dependence of $\eta$ for two different $\phi$ at $\dot\gamma=10^{-3}$. 
In the present range of $H$, $\eta$ is significantly smaller for larger $H$.  
(b) The $\phi$-dependent viscosity for active and passive rod-like particles at $H=128$ and  $\dot\gamma=10^{-3}$. Our preliminary results for passive ($F_A=0$) and active-puller ($F_A=-20$) cases show that $\eta$ increases with increasing $\phi$; the viscosity enhancement is much stronger in the active-puller case. In both (a) and (b), the dashed lines indicate the solvent viscosity $\eta_s(=1)$. 
}
\label{Fig2}
\end{figure}

First, in Fig. 2, we show the viscosity $\eta$ for various conditions. In this study, the viscosity is defined as 
\begin{eqnarray}
\eta=\dfrac{1}{{\dot \gamma}L^2}\int {\rm d}x{\rm d}y\langle \Sigma_{xz}(x,y,\pm H/2)\rangle, 
\end{eqnarray} 
where $\Sigma_{xz}(x,y,\pm H/2)$ is the $xz$ component of the stress tensor at the walls and $\langle \cdots \rangle$ hereafter denotes taking the time average in a steady state. Here, the solvent viscosity is scaled to be $1$. At a relatively low shear rate $\dot\gamma$, we find that $\eta$ takes lower values than the solvent viscosity, which qualitatively agrees with the experimental results \cite{Sokolov,Gachelin,Lopez,Liu,PNAS2020}. 
This behavior strongly depends on $H$ and the volume fraction of the swimmers defined as $\phi=N{\mathcal V}^{(b)}/L^2H$ with ${\mathcal V}^{(b)}$ being the volume of the body part. 
The viscosity can be divided into three parts: $\eta=\eta_{s}+\Delta\eta_{p}+\Delta\eta_{a}$, where $\eta_{s}(=1$ in this study) is the solvent viscosity, $\Delta\eta_{p}$ and $\Delta\eta_{a}$ are the passive and active contributions, respectively (see Appendix A for more details). 
In the present framework, $\Delta\eta_{a}$ is given as 
\begin{eqnarray}
\Delta\eta_{a} = -\dfrac{1}{{\dot\gamma}L^2 H} F_A \ell_0\sum_{\alpha=1}^N  \langle \hat{n}_{\alpha,x}(t) \hat{n}_{\alpha,z}(t)\rangle ,  \label{active_viscosity}
\end{eqnarray}
where ${\hat{\mbox{\boldmath$n$}}}_\alpha(t)$ is the unit vector representing the $\alpha$-th swimmer's  orientation at time $t$, and ${\hat{n}}_{\alpha,x}$ and ${\hat{n}}_{\alpha,z}$ are its $x$ and $z$ components, respectively. Essentially identical expressions of Eq. (\ref{active_viscosity}) were previously derived (see Refs. \cite{Haines,Saintillan1,Review3} for example). From Eq. (\ref{active_viscosity}), when swimmers tend to align along the extension direction of the flow field ($\langle \hat{n}_{\alpha,x} \hat{n}_{\alpha,z}\rangle>0$), $\Delta\eta_{a}<0$. Since the contribution of $\Delta \eta_{p}$ to $\eta$ is positive in general, a significant decrease in the viscosity occurs from the negative $\Delta \eta_a$.

To further explore what swimming states are involved in the viscosity reduction, 
we investigate the following steady-state quantities:  
${\rho}(z)=\sum_{\alpha=1}^{N}  \langle \delta[{\mbox{\boldmath$r$}}-{\mbox{\boldmath$R$}_\alpha}(t)] \rangle$, ${\mbox{\boldmath$p$}}(z)=\sum_{\alpha=1}^{N} \langle {\hat{\mbox{\boldmath$n$}}}_\alpha (t) \delta[{\mbox{\boldmath$r$}}-{\mbox{\boldmath$R$}_\alpha}(t)]\rangle$, and ${\stackrel{\leftrightarrow}{\mbox{\boldmath$W$}}}(z)=\sum_{\alpha=1}^{N}\langle [{\hat{\mbox{\boldmath$n$}}}_\alpha(t){\hat{\mbox{\boldmath$n$}}}_\alpha(t)-{\stackrel{\leftrightarrow}{\mbox{\boldmath$\delta$}}}/3] \delta[{\mbox{\boldmath$r$}}-{\mbox{\boldmath$R$}_\alpha}(t)]\rangle $. Here, ${\mbox{\boldmath$R$}_\alpha}(t)$ is the center of the force dipole ($\ne$ the center-of-mass position), $\rho(z)$ is the denisty, and ${\mbox{\boldmath$p$}}(z)/\rho(z)$ and ${\stackrel{\leftrightarrow}{\mbox{\boldmath$W$}}}(z)/\rho(z)$ represent the polarization vector and the nematic order parameter tensor, respectively \cite{Review2}.  
These quantities, which depend only on $z$ at steady state, are shown in Figs. 3(a)-(h) for various conditions. In Figs. 3(a) and (b), $\rho(z)$ has significant peaks near the boundary walls, and otherwise, it is almost constant, indicating that the walls attract swimmers. 
Such behaviors were already reported and discussed in the literature (for example, Refs. \cite{Berke,Ji-Tang,Review1,LaugaB,Figueroa-Morales,Bianchi,Denissenko,Ezhilan,Ezhilan-Saintillan,Yan-Brady}). 
 In the present model, without thermal fluctuations, when placing one swimmer near the wall, it continues to swim along the wall, which suggests that the force-dipole prescribed to the swimmer contributes to the wall attraction \cite{Berke}.
However, in the many-swimmer case, significant disturbances are induced by interactions among the swimmers. Such disturbances produce an outgoing flux from the wall to the bulk region. Meanwhile, self-propulsive motions give incoming flux to the wall from the bulk. Competition between these two flux terms should determine the amount of accumulation of swimmers at the walls \cite{Ezhilan-Saintillan,Yan-Brady}. 

Figures 3(c)-(f) show ${\mbox{\boldmath$p$}}(z)/{\rho}(z)$. Due to the flow and geometrical symmetries, $p_{y}(z)=0$ for all $z$. 
For $\phi\gtrsim 0.03$, swimmers trapped at the walls tilt their ``heads'' to the walls \cite{Vigeant,Spagnolie_Lauga,Sipos}. Moreover, the tilting angle is greater for larger  
$\phi$, which may be caused by HIs among the swimmers on the wall. These issues will be studied elsewhere.  

In terms of the viscosity reduction, among Figs. 3(a)-(h), of particular interest are (g) and (h), exhibiting $W_{xz}(z)/\rho(z)$. At $\phi=0.01$ and $H=128$, where the viscosity reduction is absent (see Fig. 2), $W_{xz}(z)\lesssim 0$ as a whole. 
For larger $\phi$ and $H$, in contrast, $W_{xz}(z)>0$ 
for all $z$. 
Within the present range of $\phi$ and $H$, by increasing these parameters, the upward convex form of $W_{xz}(z)/\rho(z)$ tends to grow. 
Equation (\ref{active_viscosity}) is rewritten as 
\begin{eqnarray}
\Delta \eta_a = -\dfrac{1}{{\dot\gamma} H}F_A \ell_0\int_{-H/2}^{H/2} {\rm d}z W_{xz}(z),  \label{active_viscosity2}
\end{eqnarray}
through which $W_{xz}(z)$ is directly related to the viscosity reduction.

\begin{figure}[hbt] 
\includegraphics[width=8.7cm]{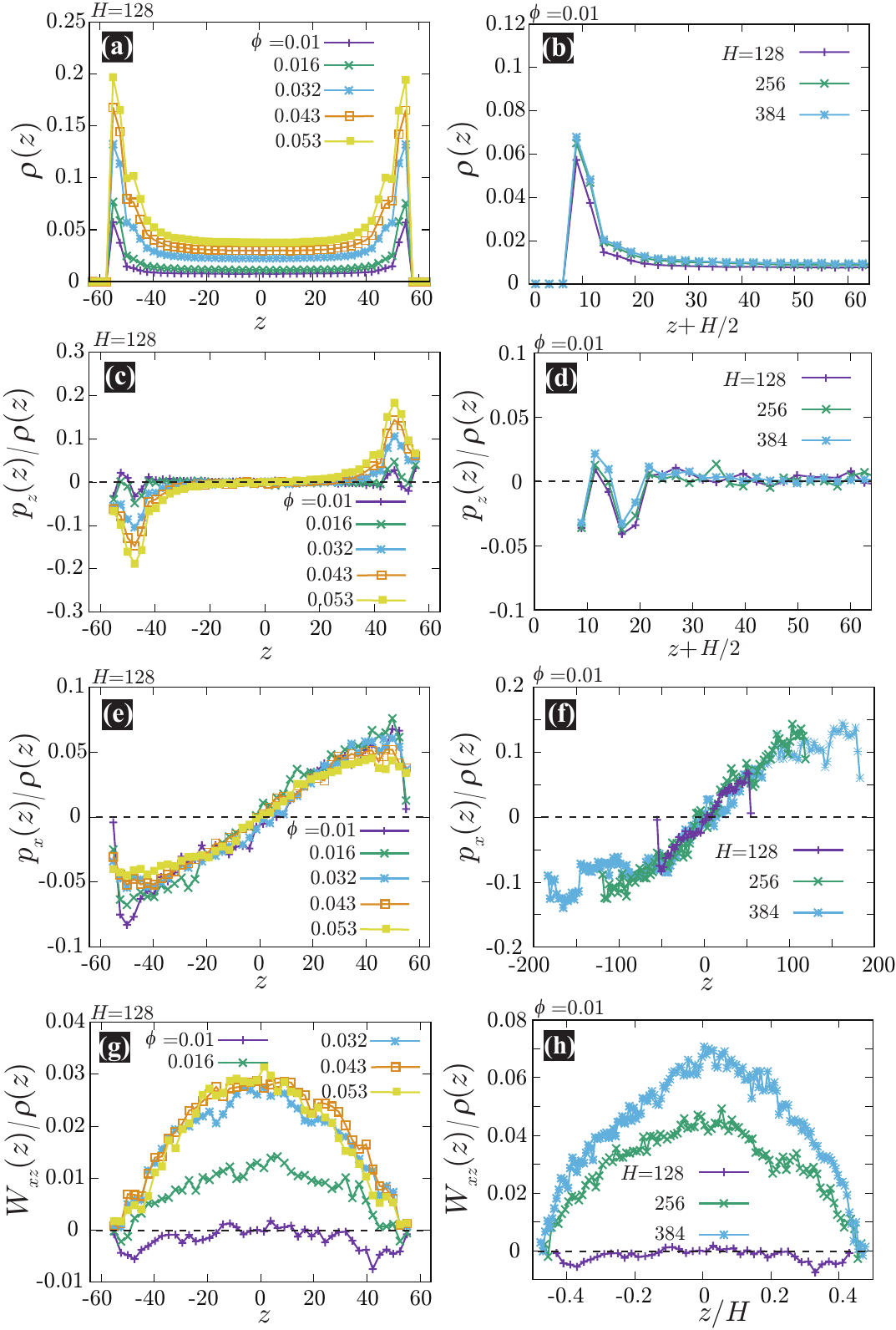}
\caption{(Color online) 
Subfigures (a), (c), (e), and (g) represent $\rho(z)$, $p_{z}(z)/\rho(z)$, $p_{x}(z)/\rho(z)$, and $W_{xz}(z)/\rho(z)$, respectively, for various $\phi$ at $H=128$ and $\dot\gamma=10^{-3}$, while the results for various $H$ at $\phi=0.01$ and $\dot\gamma=10^{-3}$ are shown in (b), (d), (f), and (h).}
\label{Fig3}
\end{figure}

\begin{figure}[hbt] 
\includegraphics[width=8.7cm]{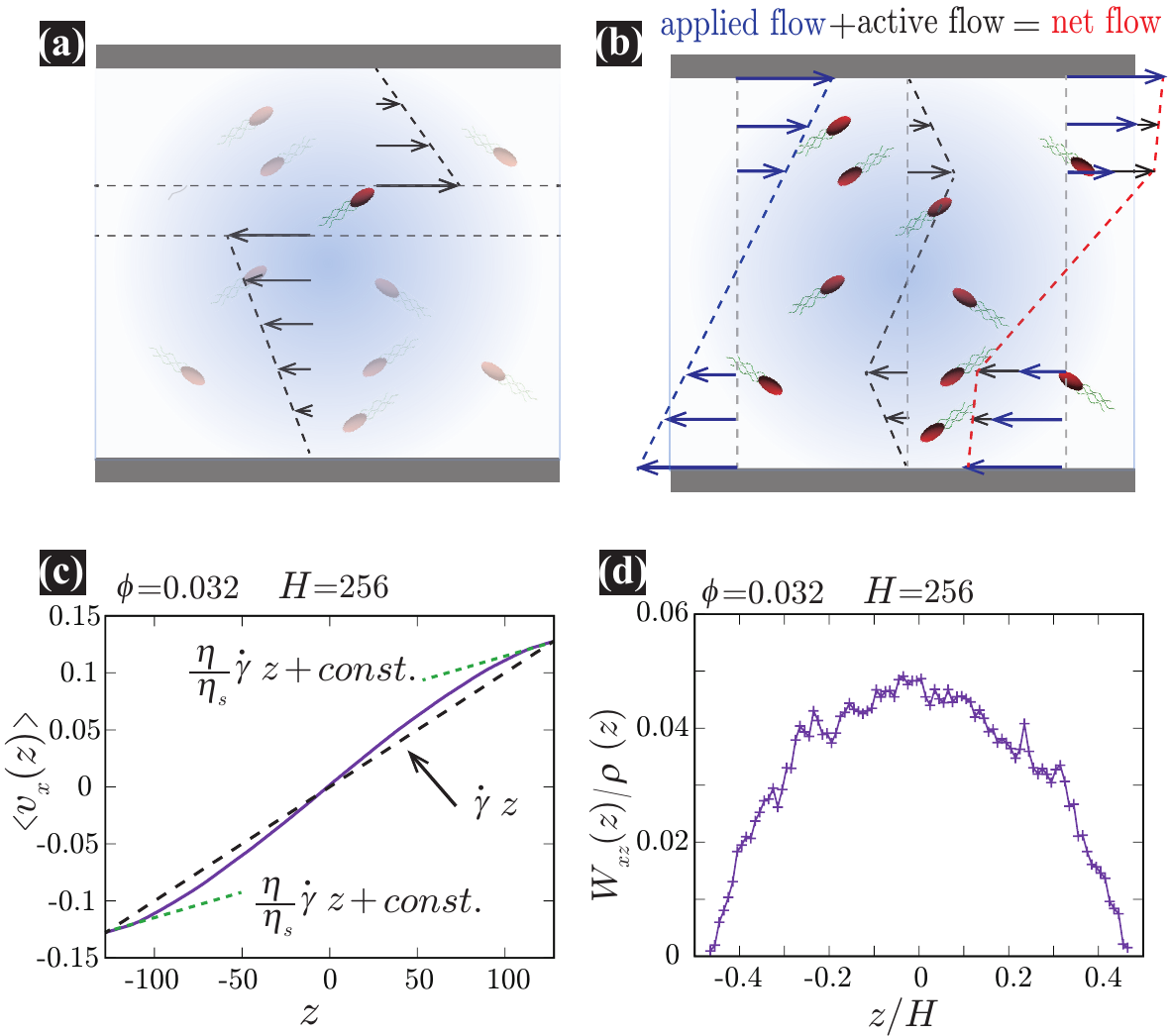}
\caption{(Color online) 
(a) For a swimmer with ${\hat n}_{\alpha,x} {\hat n}_{\alpha,z}>0$, the surrounding velocity gradient is intensified, while away from the swimmer, the opposite occurs. (b) The net flow is determined by a superposition of the individual swimmers' contributions. (c) The $x$ component of the velocity field averaged over the steady state at $\phi=0.032$ and $H=256$. At the walls, the shear rate is lower than $\dot\gamma$ as ${\dot\gamma}_w=(\eta/\eta_s){\dot \gamma}<{\dot\gamma}$, in exchange for a greater shear rate in the interior region. (d) $W_{xz}(z)/\rho(z)$ at the same condition as (c). 
}
\label{Fig4}
\end{figure}

The reduced viscosity immediately indicates the reduced shear rate at the walls.  Here, we briefly review this behavior. For a swimmer with ${\hat n}_{\alpha,x} {\hat n}_{\alpha,z}>0$, the active-force reinforces the applied flow. More specifically, in a small region including the swimmer, the velocity gradient is intensified, while in outer regions, the opposite happens. A superposition of such contributions gives the net effects on the mean flow, and we observe a lower shear rate at the walls, ${\dot\gamma}_w$, in exchange for a greater shear rate in the interior region. These situations are schematically illustrated in Figs. \ref{Fig4} (a) and (b).  Consequently, as shown in Fig. \ref{Fig4}(c), the shear stress required to maintain the applied shear rate $\dot\gamma=V/H$ is reduced \cite{Hatwalne}, and the viscosity is given by 
\begin{eqnarray}
\eta =\eta_s \dfrac{\dot\gamma_w}{\dot\gamma}, 
\end{eqnarray}
which is smaller than $\eta_s$ for $\dot\gamma_w<\dot\gamma$. Such a modulation of the velocity field accompanying with the viscosity reduction was certainly observed in experiments of {\it E. coli} suspensions \cite{PNAS2020}. 
Notably, in contrast to active (pusher) suspensions, dispersed particles in the usual passive system suppress the velocity gradient in the interior region, and the observed viscosity is increased.

\subsection{Key role of hydrodynamic interactions in determining the global swimming states}

\begin{figure}[hbt] 
\includegraphics[width=7.0cm]{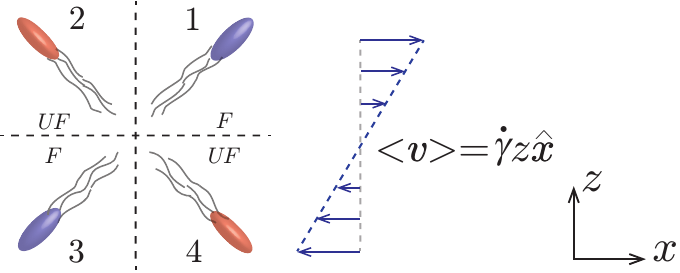}
\caption{(Color online) Fourfold classification of (one-swimmer) swimming states. These four states are further classified into the {\it favorable} $F$- and the {\it unfavorable} $UF$-states.  
}
\label{Fig5}
\end{figure}

Then, we investigate how such observed global steady states are realized. To this end, taking the flow and geometrical symmetries into account, it is useful to classify one-swimmer states into the following four states (see Fig. \ref{Fig5}): 
state 1, $({\hat n}_{\alpha,x}>0,{\hat n}_{\alpha,z}>0)$, 
state 2, $({\hat n}_{\alpha,x}<0,{\hat n}_{\alpha,z}>0)$, 
state 3, $({\hat n}_{\alpha,x}<0,{\hat n}_{\alpha,z}<0)$, 
and state 4, $({\hat n}_{\alpha,x}>0,{\hat n}_{\alpha,z}<0)$. 
States 1(2) and 3(4) are equivalent; that is, they can be converted to each other by simply rotating the coordinate frame about the $y$-axis by $\pi$. Figures 3(g) and (h) with Eq. (\ref{active_viscosity2}) indicate that, when $\Delta \eta_a<0$, states 1 and 3 with ${\hat n}_{\alpha,x}{\hat n}_{\alpha,z}>0$ are realized more favorably than states 2 and 4 with ${\hat n}_{\alpha,x}{\hat n}_{\alpha,z}<0$. Thus, we may further classify states 1 and 3 into the {\it favorable} $F$-state and states 2 and 4 into the {\it unfavorable} $UF$-state. 

Now, the question is why states 1 and 3, which contribute to $W_{xz}>0$, are more favorable than states 2 and 4. In our simulations, thermal effects are absent, and excluded volume effects are almost irrelevant because the suspensions mainly considered here are dilute. 
Instead, hydrodynamic effects are expected to determine the overall swimming states.  We expect swimmer's motions to be largely disturbed by HIs even without approaching the contact distance to other swimmers; we regard such events as hydrodynamic collisions. We support this perspective by analyzing the transition probabilities between the swimming states: we pick up a pair of swimmers whose separation at time $t$ is less than a certain close distance $d_0$; then, the transition probabilities are determined by comparing their states at $t-\Delta t$ and $t+\Delta t$. 
In this study, we set $d_0=0.7\ell_0$ and $\Delta t=1.25\tau_0$. Here, $\tau_0=\ell_0/v_s$ is the time to travel the distance of the swimmer size ($\sim\ell_0$), with $v_s$ being the average swimming speed. In the present range of $\phi$, the average distance between neighboring swimmers, $l_N=(\phi/{\mathcal V}^{(b)})^{-1/3}$ is 2 $\sim$ 3 times larger than $d_0$. 
Although quantitative evaluations of the transition probabilities significantly depend on $d_0$ and $\Delta t$, the qualitative discussion presented below is not affected as far as $d_0$ and $\Delta t$ are sufficiently smaller than $l_N$ and $l_N/v_s$, respectively. In Appendix C, we discuss how the present definition can capture hydrodynamic collisions with the settings of $d_0$ and $\Delta t$.

\begin{table}[th]
\caption{\label{tab:transition} State probabilities $P_{\mu}$ and the transition probabilities ${\mathcal W}_{\mu|\nu}$ for various conditions}
\begin{ruledtabular}
\begin{tabular}{lllllll}
$(10^2\phi, H,\dot\gamma)$ & $P_F$ & $P_{UF}$ & ${\mathcal W}_{F|F}$ & ${\mathcal W}_{F| UF}$ &  ${\mathcal W}_{UF | F}$ &  ${\mathcal W}_{UF | UF}$\\
  $(1,128,10^{-3})$ &  0.51 & 0.49 & 0.69 & 0.31 & 0.39 & 0.61\\
  $(1,384,10^{-3})$ &  0.6 & 0.4 & 0.67 & 0.33 & 0.44 & 0.56 \\
  $(3.2,128,10^{-3})$ &  0.54 & 0.46 & 0.63 & 0.37 & 0.44 & 0.56 \\
  $(3.2,128,3\times10^{-4})$ &  0.51 & 0.49 & 0.62 & 0.38 & 0.41 & 0.59 \\
  $(3.2,128,0)$ &  0.51 & 0.49 & 0.60 & 0.40 & 0.41 & 0.59 \\
\end{tabular}
\end{ruledtabular}
\end{table}

Table I shows the numerically obtained probability of the $\mu$-state, $P_{\mu}$, and the transition probability from the $\mu$- to $\nu$-states, ${\mathcal W}_{\mu|\nu}$, ($\mu,\nu$=$F, UF$) for various conditions.  Here, $P_{\mu}$ and ${\mathcal W}_{\mu|\nu}$ are calculated for swimmers in the region $-0.3 \le z/H \le 0.3$. At ${\dot\gamma}=0$, because the $F$- and $UF$-states are not distinguished, ${\mathcal W}_{F| UF}\cong {\mathcal W}_{UF| F}$, and $P_F\cong P_{UF}$. However, for ${\dot\gamma}\ne 0$, we find that ${\mathcal W}_{F| UF}$ is significantly smaller than ${\mathcal W}_{UF| F}$. As $H$ and $\phi$ increase (in the dilute regime), the population of the $F$-state swimmers with ${\hat n}_{\alpha,x}{\hat n}_{\alpha,z}>0$ increases, indicating that an increase in the collision frequency or time further promotes transitions. 

For swimmers trapped at the walls, the hydrodynamic torques arising from the applied flow weakly align them along the flow direction because their heads are slightly tilted against the walls. Thus, for trapped swimmers, the population of the $F$-state is slightly larger than that of the $UF$-state. 
After longer-term traps, swimmers leave from the bottom (top) wall by raising (dropping) their heads, which changes their states ($F\rightleftarrows UF$). 
Reflecting such conditions, for swimmers just after leaving the walls, the population of the $UF$-state is slightly larger than that of the $F$-state (not shown here).
As swimmers move inward from the boundary walls, transitions from the $UF$- to $F$-states are gradually promoted by collisions. Due to the geometrical symmetry, the population of the $F$-state is maximized at $z=0$, leading to the upward convex form of $W_{xz}(z)/\rho(z)$. 
In an ideal bulk system or a system with periodic boundary conditions without walls, a detailed balance between the $F$- and $UF$-states, $P_F {\mathcal W}_{F|UF}=P_{UF}{\mathcal W}_{UF| F}$, should be realized. Such a detailed balance may nearly hold at larger $H$ and $\phi$ in the present system, but that was not investigated in detail.  

\begin{table}[htb]
\caption{\label{tab:transition}  Transition probability from states 1 to $\mu$ through a hydrodynamic collision with another swimmer in state $\nu$, $w_{1 | \mu}^{(\nu)}$, at $\phi=0.01$, $H=384$, and $\dot\gamma=10^{-3}$.  }
\begin{ruledtabular}
\begin{tabular}{lllll}
 $\nu$ & $w_{1|1}^{(\nu)}$ & $w_{1|2}^{(\nu)}$ & $w_{1|3}^{(\nu)}$ & $w_{1|4}^{(\nu)}$ \\
  1 & 0.75  & 0.05  & 0.01 & 0.19 \\
  2 & 0.67  & 0.20  & 0.03 &  0.10 \\  
  3 & 0.61  & 0.10  & 0.07 & 0.22 \\
  4 & 0.52  & 0.04  & 0.02 & 0.42 \\

\end{tabular}
\end{ruledtabular}
\end{table}
\begin{table}[htb]
\caption{\label{tab:transition}   Transition probability from states 2 to $\mu$ through a collision with another swimmer in state $\nu$, $w_{2 | \mu}^{(\nu)}$, at $\phi=0.01$, $H=384$, and $\dot\gamma=10^{-3}$.   }
\begin{ruledtabular}
\begin{tabular}{lllll}
 $\nu$ & $w_{2| 1}^{(\nu)}$ & $w_{2| 2}^{(\nu)}$ & $w_{2| 3}^{(\nu)}$ & $w_{2| 4}^{(\nu)}$ \\
  1 & 0.37  & 0.51 & 0.06 & 0.05 \\
  2 & 0.19  & 0.66 & 0.15 & 0.00 \\
  3 & 0.14  & 0.46 & 0.37&  0.03 \\
  4 & 0.25  & 0.51  & 0.19 & 0.05  \\
\end{tabular}
\end{ruledtabular}
\end{table}

Tables II and III show the numerically obtained transition probabilities of a swimmer in states 1 and 2 before a collision, respectively, at $\phi=0.01$, $H=384$, and $\dot\gamma=10^{-3}$. 
Here, $w_{\lambda | \mu}^{(\nu)}$ represents the transition probability from states $\lambda$ to $\mu$ through a collision with another swimmer in state $\nu$.  Note that similar results are obtained at different parameters where negative $\Delta\eta_a$ is obtained. We find significant differences between $w_{1 | \mu}^{(\nu)}$ and $w_{2 | \mu}^{(\nu)}$. For both cases the majorities are $w_{\mu|\mu}^{(\nu)}$, whereas a swimmer in state 1 is more likely to retain its state unchanged by a collision than one in state 2. 

We can understand the role of HIs in the elementary processes of these state transitions through the following phenomenological arguments, which are separately provided for different cases.   

\subsubsection{Hydrodynamic collisions between two swimmers in states 1 and 2, and 2 and 3}

\begin{figure}[hbt] 
\includegraphics[width=8.7cm]{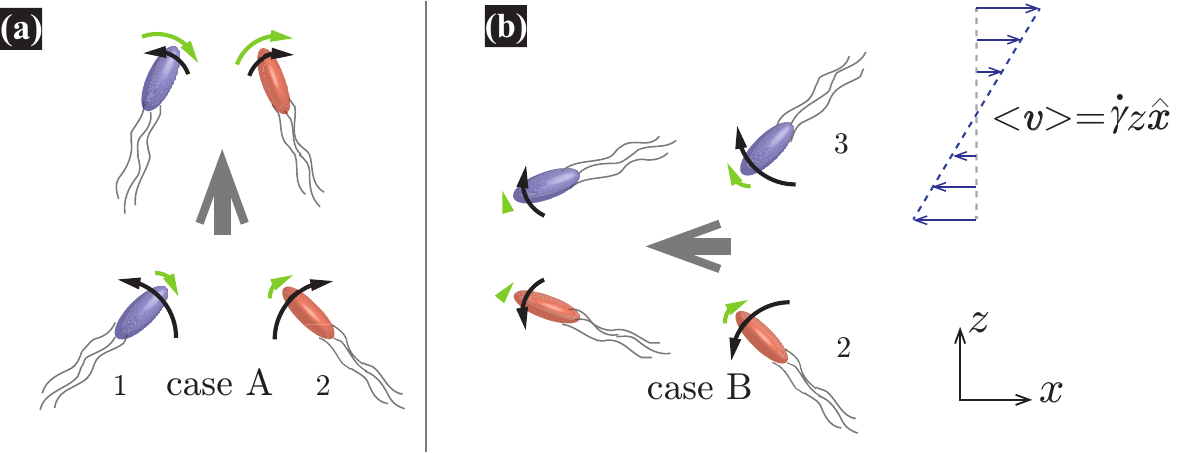}
\caption{(Color online) 
Schematic illustrations for two approaching swimmers in the $F$- and $UF$-states: case (A) where the two swimmers are in states 1 and 2 (a), and case (B) where those are in states 2 and 3 (b). The torques due to HIs and the shear flow are described by the black and green curved arrows, respectively. 
}
\label{Fig6}
\end{figure}

Let us first consider that two swimmers in the $F$- and $UF$-states are approaching. There are essentially two different cases: case (A) where swimmers in states 1 and 2  (equivalently, 3 and 4) are approaching, and case (B) where swimmers in states 3 and 2 (equivalently, 1 and 4) are approaching. 

For case (A), as schematically shown in Fig. \ref{Fig6}(a), HIs tend to rotate the swimmers in opposite directions \cite{LaugaB}, while the externally applied shear flow rotates them in the same direction. As the swimming directions become parallel to each other and perpendicular to the flow direction, the torques due to HIs grow weaker, but those arising from the shear flow grow stronger. 
Furthermore, once two swimmers move nearly side by side (Fig. \ref{Fig6}(a)), a hydrodynamic attraction acts on them, which may make a swimmer in state 1 drag one in state 2 into eventually moving in the same direction in the collision process. 
These hydrodynamic effects are expected to promote the transition from states 2 to 1,  responsible for $w_{1 | 2}^{(2)}< w_{2 | 1}^{(1)}$ and $w_{1 | 1}^{(2)} > w_{2 | 2}^{(1)}$. 

In contrast, in case (B), due to similar asymmetry in the net torques, $w_{3 | 2}^{(2)}$ (not shown here but equivalent to $w_{1 | 4}^{(4)}$) is slightly larger than $w_{2 | 3}^{(3)}$. 
In this case, the torques both due to HIs and the shear flow are reduced as their swimming directions become parallel along with the flow direction (see Fig. \ref{Fig6}(b) for a schematic), and therefore, the difference between $w_{3 | 2}^{(2)}$ ($\cong w_{1 | 4}^{(4)}$) and $w_{2 | 3}^{(3)}$ is less notable than that between $w_{1 | 2}^{(2)}$ and $w_{2 | 1}^{(1)}$: namely, hydrodynamic collisions of case (A) predominantly contribute to the transition from the $UF$ to $F$ states, whereas those of case (B) are marginal. 

\subsubsection{Hydrodynamic collisions between two swimmers in the same states} 

\begin{figure}[hbt] 
\includegraphics[width=8.7cm]{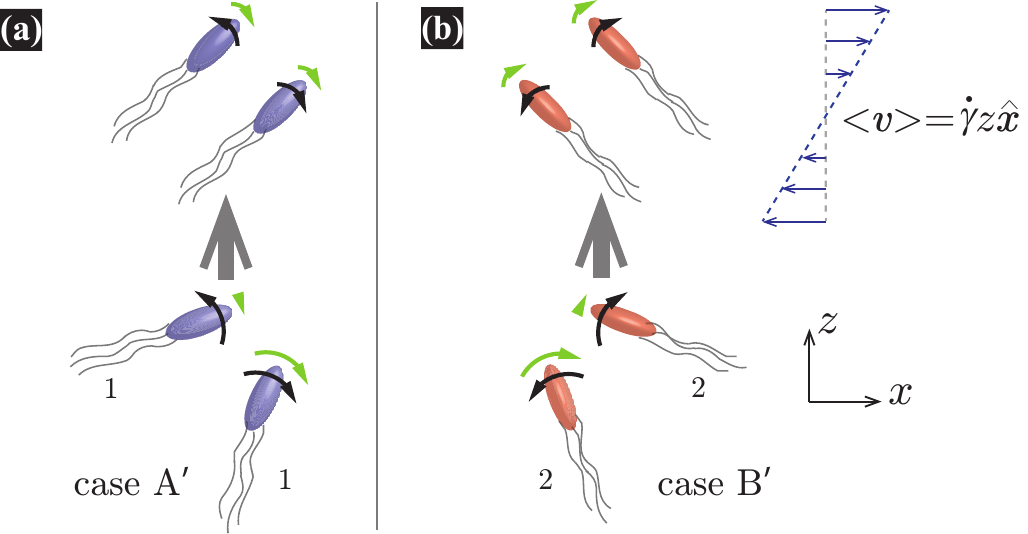}
\caption{(Color online) 
Schematic illustrations of the two cases: case (A') where the two swimmers are both in state 1 (a) and case (B') where those are both in state 2 (b). The torques due to HIs and the shear flow are described by the black and green curved arrows, respectively. 
}
\label{Fig7}
\end{figure}

Here, we consider the following  two cases: case (A') where two swimmers are both in state 1 (equivalently, both in state 3), and case (B') where those are both in state 2 (equivalently, both in state 4). 
For these cases, schematics are shown in Figs. \ref{Fig7}(a) and (b).  

For both cases (A') and (B'), the torques caused by HIs rotate the swimmers in opposite directions and grow weaker as the swimmers become parallel to each other.  
On the other hand, for the torques caused by the shear flow, in case (A'), as the collision proceeds, the torque resisting the transition to state 2 grows stronger, whereas the other torque, which helps the transition to state 4, grows weaker. 
In (B'), the opposite occurs: one torque due to the shear flow promoting the transition to state 1 grows stronger, while the other one resisting the transition to state 3 grows weaker. 
The difference in how the torques contribute to the transition is expected to be responsible for the measured difference in the transition probabilities.   
That is, as shown in Tables II and III, $w_{1| 2}^{(1)}<w_{2| 1}^{(2)}$ and $w_{2| 3}^{(2)} \lesssim w_{1| 4}^{(1)}$, resulting in $w_{1| 1}^{(1)}>w_{2| 2}^{(2)}$. 

\subsubsection{Hydrodynamic collisions between two swimmers in states 1 and 3, and 2 and 4}

\begin{figure}[hbt] 
\includegraphics[width=8.7cm]{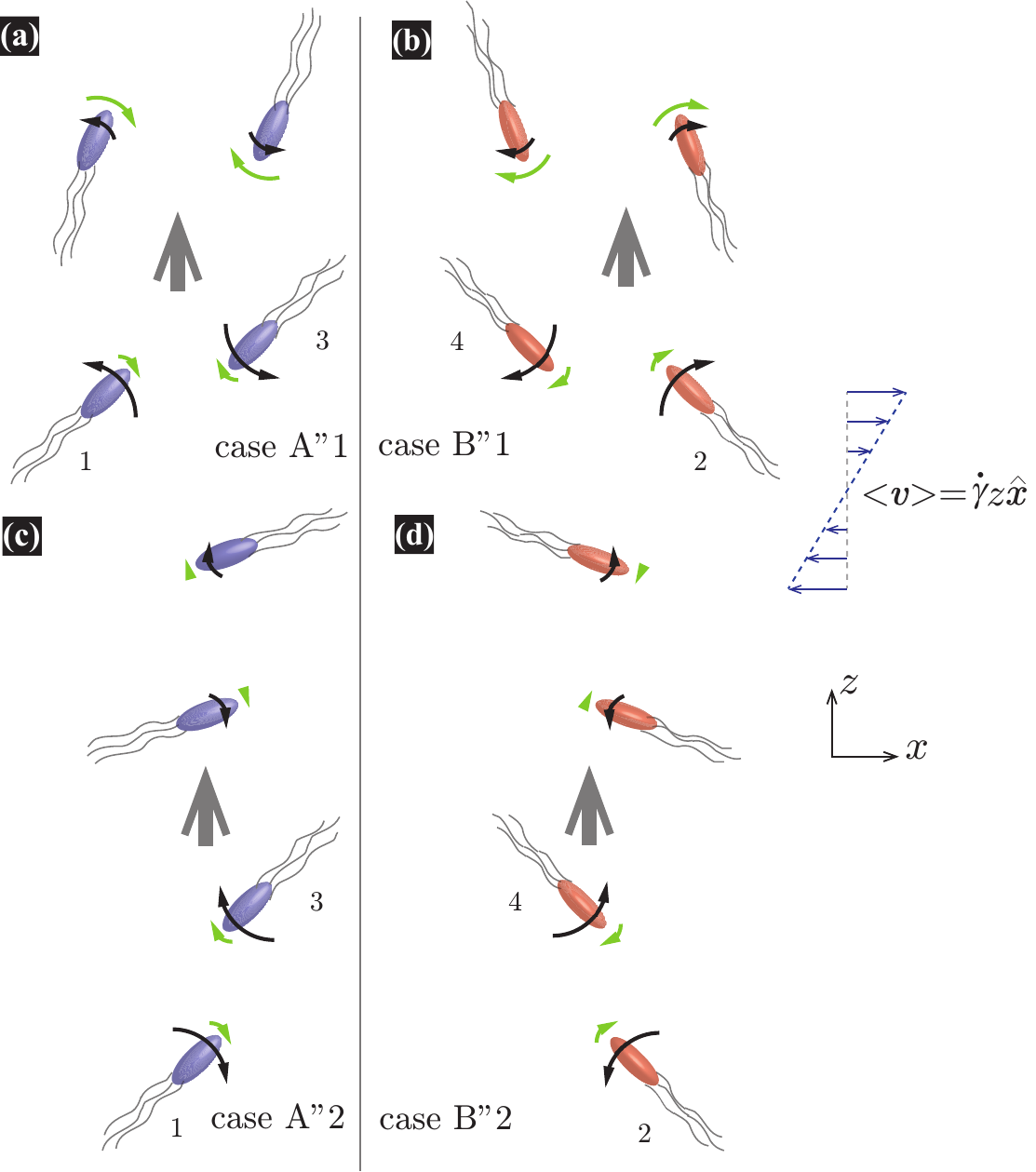}
\caption{(Color online) 
Schematic illustrations of cases (A''1) and (A''2), shown in (a) and (c), respectively, where the two swimmers are in states 1 and 3, and those of cases (B''1) and (B''2), shown in (b) and (d), respectively, where the two swimmers are in states 2 and 4. 
The torques due to HIs and the shear flow are described by the black and green curved arrows, respectively. 
}
\label{Fig8}
\end{figure}

When two approaching swimmers are in states 1 and 3, there are essentially two different cases, (A''1) and (A''2),  which are illustrated in Figs. \ref{Fig8}(a) and (c), respectively.  
As torques arising from HIs rotate the two swimmers in opposite directions, the swimming directions become perpendicular and parallel to the flow in cases (A''1) and (A''2), respectively. In case (A''1), the torques due to the shear flow, which prevent the swimmers from changing from the $F$- to $UF$-states, grow stronger. On the other hand, in case (A''2), such torques promoting the changes to the $UF$-state grow weaker. 

When swimmers in states 2 and 4 are approaching, there are two different cases, (B''1) and (B''2), as schematically shown in Figs. \ref{Fig8}(b) and (d), respectively.   
With similar arguments, in case (B''1), the torques due to the shear flow, which promote the transition to the $F$-state, grow stronger, while, in case (B''2), those preventing the changes to the $F$-state grow weaker.  

These differences may result in the difference between $w_{1| \mu}^{(3)}$ and $w_{2| \mu}^{(4)}$. That is, when two swimmers in states 1 and 3 are approaching each other, the swimmers tend to remain their states unchanged more than when they are in states 2 and 4: $w_{1| 2}^{(3)}<w_{2| 1}^{(4)}$ and $w_{1| 4}^{(3)} \gtrsim w_{2| 3}^{(4)}$, resulting in $w_{1| 1}^{(3)}>w_{2| 2}^{(4)}$. Note that hydrodynamic collisions of cases (A''1) and (B''1) predominantly contribute to the transition from the $UF$ to $F$ states, whereas those of cases (A''2) and (B''2) are marginal.

The realistic collision processes are more complicated; thus, the present arguments are oversimplified. However, they qualitatively explains why HIs systematically promote the transition from the $UF$ to $F$ states.

\section{Discussion and Concluding remarks}

It has been known that rod-like particles in a shear flow tend to orientate to the extension direction due to an interplay between flow and rotational diffusivities \cite{Hinch_Leal1,Hinch_Leal2}: for a rod-like particle, although the torque due to shear flow becomes unidirectional and stronger as its orientation becomes perpendicular to the flow direction, the torque due to thermal rotational diffusivities is bidirectional and does not depend on the rod orientation. By considering such an effect, an explanation for the viscosity reduction was provided \cite{Haines,Saintillan1}. 
Moreover, it was proposed that the activity-induced HIs provide a source of random rotations in addition to thermal fluctuations and tumbling \cite{Ryan}. 
The present study further illuminates the role of HIs: even starting from a random state, our results suggest that steady global states where the swimmers are weakly aligned along the extension axis may form as self-organization by repeated hydrodynamic collisions. A study of this issue would be an interesting task for future studies. 

In typical microorganisms systems, the propulsive forces are sufficiently strong that hydrodynamic effects may dominate over thermal fluctuations.  
Below, we validate this condition by considering a typical experimental situation \cite{PNAS2020}: an {\it E. coli} suspension at a volume fraction $\phi(={\mathcal V}^{(b)}/l_N^3)=$0.01 at room temperature ($\sim 300$K), for which the average separation distance is $l_N \sim 5 \mu$m and the thermal rotational diffusion coefficient is $D_T\lesssim$1s$^{-1}$. 
Hereafter, we assume that the swimming speed is $v_s\sim 10\mu$m/s, the magnitude of the force dipole is $P\sim10^{-18}$N$\cdot$m, the cell size is $\ell_0\sim 1\mu$m, the cell volume is ${\mathcal V}^{(b)}\sim 1 \mu$m$^3$, and the solvent viscosity is $\eta_s\sim 10^{-3}$Pa$\cdot$s. 
For a duration $\sim 1/D_T\gtrsim1$s, a swimmer may at least once approach another swimmer closer than $l_c \sim 2\mu$m, estimated by $\pi l_c^2 v_s\times (1/D_T)\times (1/l_N)^3\sim 1$.  
The magnitude of the rotational flow field, $\omega$, induced at a distance $r$ from a swimmer is approximately given as $\omega\sim 0.1 P/(\eta_{s} r^3)$ \cite{LaugaB}. Therefore, at $r=l_N$, $\omega\sim 1$s$^{-1}$, while at $r=l_c$, $\omega\sim 10$s$^{-1}$. 
By such a hydrodynamic ``collision'' process, which lasts for approximately $\ell_0/v_s\sim 0.1$s,  swimming motions can be largely affected more than by thermal fluctuations. 
In other words, reorientation due to HIs may be a faster process than thermal rotational diffusion. 

In this study, we have explored a mechanism of the anomalous rheology of active suspensions, focusing on the role of HIs. Before closing, we present the following remarks. 
(1) Our pusher model is transformed into a puller model by simply changing the sign of the active forces. Our preliminary results shown in Fig. 2(b) suggest that the viscosity of the puller model is increased more than that in the passive systems, which agrees with experimental observations for motile and immotile puller bacterial suspensions \cite{Rafai}. 
(2) In Ref. \cite{Liu}, under Poiseuille flow, lower viscosity is observed for smaller separation between the walls. This contrasts with the present result, where the viscosity reduction is enhanced by increasing $H$ under simple shear. In addition, the viscosity reduction occurs at larger shear rates than those of the experiments of Refs. \cite{Lopez,PNAS2020}. These differences may be attributed to the difference in the flow geometry.  
We plan to investigate these issues further elsewhere.

\begin{acknowledgments}

This work was supported by KAKENHI (Grants No. 26103507, 25000002, and 20H05619) and the special fund of Institute of the Industrial Science, The University of Tokyo. 

\end{acknowledgments}

\appendix 

\section{simulation method}
In our simulations, we use the smoothed-profile method (SPM) \cite{SPM,SPM2,SPM3} to accomodate many-body hydrodynamic interactions (HIs) among the constituent swimmers. 
In Ref. \cite{SPM3}, it is found that the SPM can quantitatively reproduce far-field and intermediate-field aspects of HIs, whereas the near-field HIs are slightly underestimated at closer distances. Furthermore, like many other methods, the SPM cannot also resolve the singular lubrication forces. For more details of the qualitative evaluations on the SPM, please refer to Refs. \cite{SPM2,SPM3}. 

\begin{figure}[bth] 
\includegraphics[width=8.7cm]{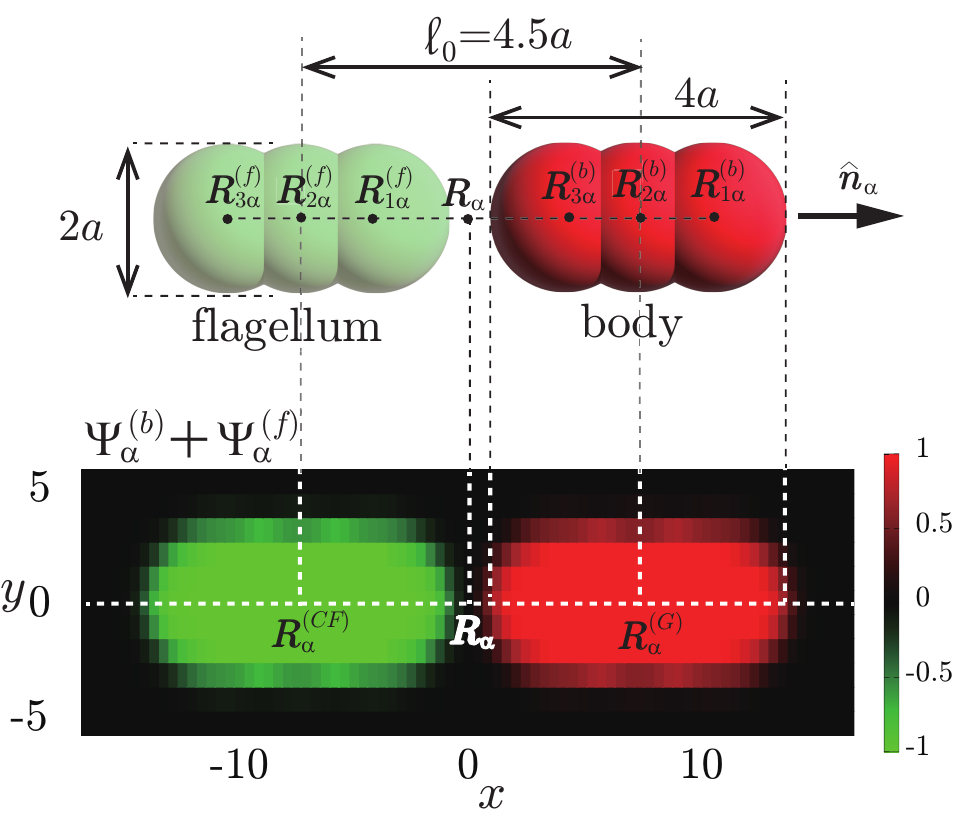}
\caption{(Color online) 
In this study, to incorporate the present model swimmer into the SPM, the body and flagellum parts are represented through the field variables, $\Psi_\alpha^{(b)}({{\mbox{\boldmath$r$}}})$ and $\Psi_\alpha^{(f)}({{\mbox{\boldmath$r$}}})$, respectively. We plot $\Psi_\alpha^{(b)}({{\mbox{\boldmath$r$}}})+\Psi_\alpha^{(f)}({{\mbox{\boldmath$r$}}})$ in the $xy$-plane, where both ${{\mbox{\boldmath$R$}}}_{\alpha}^{(G)}=(2.25R,0,0)$ and  ${{\mbox{\boldmath$R$}}}_{\alpha}^{(CF)}=(-2.25R,0,0)$ are included.  The discretized mesh size $h$ is the same as that used in practical simulations ($h=0.3125R$ and $\xi=0.5h$). Here, $\xi$ is the interface thickeness controlling the degree of smoothness of $\Psi_\alpha^{(b)}({{\mbox{\boldmath$r$}}})$ and $\Psi_\alpha^{(f)}({{\mbox{\boldmath$r$}}})$. 
}
\label{afig1}
\end{figure}

For this purpose, the body and flagellum parts described above are represented through the field variables, $\Psi_\alpha^{(b)}({{\mbox{\boldmath$r$}}})$ and $\Psi_\alpha^{(f)}({{\mbox{\boldmath$r$}}})$, respectively:  
\begin{eqnarray}
\Psi_\alpha^{(b)}({{\mbox{\boldmath$r$}}}) &=& \sum_{i=1}^{3}\psi [{{\mbox{\boldmath$r$}}}, {{\mbox{\boldmath$R$}}}_{i,\alpha}^{(b)}] ~~~~ \biggl({\rm for}~~\sum_{i=1}^{3}\psi[{{\mbox{\boldmath$r$}}}, {{\mbox{\boldmath$R$}}}_{i,\alpha}^{(b)}]
\le 1 \biggr) \nonumber \\\\ 
&=& 1 ~~~~\biggl({\rm for}~~\sum_{i=1}^{3}\psi [{{\mbox{\boldmath$r$}}}, {{\mbox{\boldmath$R$}}}_{i,\alpha}^{(b)}]>1 \biggr). \label{profileB}
\end{eqnarray}
and 
\begin{eqnarray}
\Psi_\alpha^{(f)}({{\mbox{\boldmath$r$}}}) &=& -\sum_{i=1}^{3}\psi [{{\mbox{\boldmath$r$}}}, {{\mbox{\boldmath$R$}}}_{i,\alpha}^{(f)}] ~~~~ \biggl({\rm for}~~\sum_{i=1}^{3}\psi[{{\mbox{\boldmath$r$}}}, {{\mbox{\boldmath$R$}}}_{i,\alpha}^{(f)}]
\le 1 \biggr) \nonumber \\\\ 
&=& -1 ~~~~ \biggl({\rm for}~~\sum_{i=1}^{3}\psi [{{\mbox{\boldmath$r$}}}, {{\mbox{\boldmath$R$}}}_{i,\alpha}^{(f)}]>1 \biggr).  \label{profileF}
\end{eqnarray}
In this study, we adopt the following function to $\psi$ as 
\begin{eqnarray}
\psi [{{\mbox{\boldmath$r$}}}, {{\mbox{\boldmath$R$}}}_{i,\alpha}^{(\mu)}]=\dfrac{1}{2}\biggl\{\tanh\biggl[\dfrac{1}{\xi}(R-|{{\mbox{\boldmath$r$}}}-{{\mbox{\boldmath$R$}}}_{i,\alpha}^{(\mu)}|)\biggr] +1 \biggr\},  
\end{eqnarray} 
where $\mu=b,f$, and $\xi$ is the interface thickness controlling the degree of smoothness. 
In Fig. \ref{afig1}, we show the cross section of the model swimmer described by $\Psi_\alpha^{(b)}({{\mbox{\boldmath$r$}}})$ and $\Psi_\alpha^{(f)}({{\mbox{\boldmath$r$}}})$ including both ${{\mbox{\boldmath$R$}}}_{\alpha}^{(G)}$ and  ${{\mbox{\boldmath$R$}}}_{\alpha}^{(CF)}$ in the same plane.

The working equations for the velocity field  ${{\mbox{\boldmath$v$}}}({{\mbox{\boldmath$r$}}},t)$ are given as 
\begin{eqnarray}
\rho\biggl(\dfrac{\partial}{\partial t}+ {{\mbox{\boldmath$v$}}}\cdot {{\mbox{\boldmath$\nabla$}}}\biggr){{\mbox{\boldmath$v$}}} &=&  {{\mbox{\boldmath$\nabla$}}}\cdot {\stackrel{\leftrightarrow}{\mbox{\boldmath$\Sigma$}}_{vis}}-{{\mbox{\boldmath$\nabla$}}}p+ {{\mbox{\boldmath$f$}}}_{H}+ {{\mbox{\boldmath$f$}}}_{A}^{(f)},   \nonumber \\
\label{Navier_Stokes} \\ 
{\stackrel{\leftrightarrow}{\mbox{\boldmath$\Sigma$}}_{vis}} &=&  \eta_s \bigl[{{\mbox{\boldmath$\nabla$}}} {{\mbox{\boldmath$v$}}}+ ({{\mbox{\boldmath$\nabla$}}} {{\mbox{\boldmath$v$}}})^\dagger\bigr], \label{stress_tensor} \\
 {{\mbox{\boldmath$\nabla$}}}\cdot {{\mbox{\boldmath$v$}}}&=&0. \label{incompressibility}
\end{eqnarray}
Equation (\ref{Navier_Stokes}) is the usual Navier-Stokes equation \cite{Landau_LifshitzB}. 
Here, ${\stackrel{\leftrightarrow}{\mbox{\boldmath$\Sigma$}}_{vis}}$ given in Eq. (\ref{stress_tensor}) is the viscous stress tensor with $\eta_s$ being the solvent viscosity, and the hydrostatic pressure $p$ is determined by the incompressibility condition, Eq.  (\ref{incompressibility}). 
In addition, ${{\mbox{\boldmath$f$}}}_{H}$ is the body force required to satisfy the rigid body condition, and 
${{\mbox{\boldmath$f$}}}_{A}^{(f)}$ is the active force directly exerted by the flagellum part to the fluid:  
\begin{eqnarray}
{{\mbox{\boldmath$f$}}}_{A}^{(f)}({{\mbox{\boldmath$r$}}})= \dfrac{1}{{\mathcal V}_{\alpha}^{(f)}}\sum_{\alpha=1}^{N}\Psi_\alpha^{(f)}({{\mbox{\boldmath$r$}}}){\hat{\mbox{\boldmath$n$}}}_{\alpha}F_A,  \label{active_forceF}
\end{eqnarray}
where ${\mathcal V}_{\alpha}^{(f)}=\int {\rm d}{{\mbox{\boldmath$r$}}} \Psi_\alpha^{(f)}({{\mbox{\boldmath$r$}}})$ is the volume of the flagellum part. 
In addition, the volume of the body part is give as ${\mathcal V}_{\alpha}^{(b)}=\int {\rm d}{{\mbox{\boldmath$r$}}} \Psi_\alpha^{(b)}({{\mbox{\boldmath$r$}}})$. 
In this study, because the shapes of the body and flagellum parts are assumed to be the same, ${\mathcal V}_{\alpha}^{(b)}={\mathcal V}_{\alpha}^{(f)}$.  

As described in the main text, the periodic boundary conditions are imposed in the $x$- and $y$-directions with the linear dimension $L$, and the planar top and bottom walls are placed at $z=H/2$ and $-H/2$, respectively, with $H$ being the separation distance. 
The shear flow is imposed by moving the top and bottom walls in the $x$-direction at constant velocities $V/2$ and $-V/2$, respectively, whereby the mean shear rate is given as $\dot\gamma=V/H$. 
We impose no-slip boundary conditions at the top and bottom walls: ${{\mbox{\boldmath$v$}}}(x,y,H/2)=(V/2){\hat {\mbox{\boldmath$x$}}}$ and ${{\mbox{\boldmath$v$}}}(x,y,-H/2)=-(V/2){\hat {\mbox{\boldmath$x$}}}$.

The equations of motions of the center-of-mass velocity, ${{\mbox{\boldmath$V$}}}_\alpha^{(G)}$, and the angular velocity with respect to the center-of-mass, ${{\mbox{\boldmath$\Omega$}}}_\alpha^{(G)}$,  are 
\begin{eqnarray}
M_\alpha \dfrac{d{{\mbox{\boldmath$V$}}}_\alpha^{(G)}}{dt} &=& {{\mbox{\boldmath$F$}}}_{\alpha, {H}} + {{\mbox{\boldmath$F$}}}_{\alpha, {int}}+ {{\mbox{\boldmath$F$}}}_{\alpha, {A}}^{(b)}
+ {{\mbox{\boldmath$F$}}}_{\alpha, {ex}}, \nonumber \\ \label{VG} 
\\ 
{\stackrel{\leftrightarrow}{\mbox{\boldmath$I$}}}_{\alpha}\cdot 
\dfrac{d{{\mbox{\boldmath$\Omega$}}}_\alpha^{(G)}}{dt} &=&  {{\mbox{\boldmath$N$}}}_{\alpha, {H}} + {{\mbox{\boldmath$N$}}}_{\alpha, {int}}+{{\mbox{\boldmath$N$}}}_{\alpha, {ex}},   \label{OG}
\end{eqnarray}
where
\begin{eqnarray}
M_{\alpha} = \rho{\mathcal V}_\alpha^{(b)} \label{mass} 
\end{eqnarray} 
and 
\begin{eqnarray}
{\stackrel{\leftrightarrow}{\mbox{\boldmath$I$}}}_{\alpha} &=&   \int {\rm d}{{\mbox{\boldmath$r$}}} \rho\Psi_\alpha^{(b)}({{\mbox{\boldmath$r$}}})\biggl[|\Delta{{\mbox{\boldmath$r$}}}_\alpha |^2 {\stackrel{\leftrightarrow}{\mbox{\boldmath$\delta$}}} - \Delta{{\mbox{\boldmath$r$}}}_\alpha \Delta{{\mbox{\boldmath$r$}}}_\alpha \biggr]
\end{eqnarray}
are the mass and the moment of inertia of the $\alpha$-th swimmer's body, respectively. 
Here, $\Delta{{\mbox{\boldmath$r$}}}_\alpha= {{\mbox{\boldmath$r$}}}-{{\mbox{\boldmath$R$}}}_\alpha^{(G)}$. 
In this study, the swimmer's density is assumed to be the same as the solvent density. 
In Eqs. (\ref{VG}) and (\ref{OG}),  ${{\mbox{\boldmath$F$}}}_{\alpha, {int}}^{(G)}$ and ${{\mbox{\boldmath$N$}}}_{\alpha, {int}}^{(G)}$ are the force and torque acting on the $\alpha$-th swimmer's body, respectively, due to the particle-particle and particle-wall potential interactions: 
\begin{widetext}
\begin{eqnarray}
{{\mbox{\boldmath$F$}}}_{\alpha, {int}} &=& -\sum_{\beta\ne \alpha}\sum_{i,\mu\in \alpha}\sum_{j,\nu\in \beta} \dfrac{\partial}{\partial {{\mbox{\boldmath$R$}}}_{i\alpha}^{(\mu)}}U^{\mu\nu}(|{{\mbox{\boldmath$R$}}}_{i\alpha}^{(\mu)}-{{\mbox{\boldmath$R$}}}_{j\beta}^{(\nu)}|)-\sum_{i,\mu\in \alpha}  \dfrac{\partial}{\partial {{\mbox{\boldmath$R$}}}_{i\alpha}^{(\mu)}}\biggl[W^{\mu}\biggl(|z_{i\alpha}^{(\mu)}-\frac{H}{2}|\biggr) + W^{\mu}\biggl(|z_{i\alpha}^{(\mu)}+\frac{H}{2}|\biggr)\biggr]\nonumber \\
 \\
{{\mbox{\boldmath$N$}}}_{\alpha, {int}} &=& -\sum_{\beta\ne \alpha}\sum_{i,\mu\in \alpha}\sum_{j,\nu\in \beta} ({{\mbox{\boldmath$R$}}}_{i,\alpha}^{(\mu)} - {{\mbox{\boldmath$R$}}}_{\alpha}^{(G)})   \times\dfrac{\partial}{\partial {{\mbox{\boldmath$R$}}}_{i\alpha}^{(\mu)}}U^{\mu\nu}(|{{\mbox{\boldmath$R$}}}_{i\alpha}^{(\mu)}-{{\mbox{\boldmath$R$}}}_{j\beta}^{(\nu)}|)\nonumber \\
 && -\sum_{i,\mu\in \alpha} ({{\mbox{\boldmath$R$}}}_{i,\alpha}^{(\mu)} - {{\mbox{\boldmath$R$}}}_{\alpha}^{(G)})   
 \times\dfrac{\partial}{\partial {{\mbox{\boldmath$R$}}}_{i\alpha}^{(\mu)}}\biggl[W^{\mu}\biggl(|z_{i\alpha}^{(\mu)}-\frac{H}{2}|\biggr) + W^{\mu}\biggl(|z_{i\alpha}^{(\mu)}+\frac{H}{2}|\biggr)\biggr],
\end{eqnarray}
\end{widetext}
where $i,j=1,2,3$ and $\mu,\nu=b,f$. 
Here, $U^{\mu\nu}$ is the interaction potential between two spheres which each comprise the body or the flagellum part of different swimmers, and $W^{\mu}$ is the interaction potential between such a sphere and the planar wall. 
The explicit forms of $U^{\mu\nu}$ and $W^\mu$ are provided below. In Eqs. (\ref{VG}) and (\ref{OG}), 
${{\mbox{\boldmath$F$}}}_{\alpha, {ext}}$ and ${{\mbox{\boldmath$N$}}}_{\alpha, {ext}}$ are the force and torque exerted on the $\alpha$-th swimmer due to the external field, which are absent in the present study. 
The active force acting on the body part, ${{\mbox{\boldmath$F$}}}_{\alpha, {A}}^{(b)}$, is given as 
\begin{eqnarray}
{{\mbox{\boldmath$F$}}}_{\alpha, {A}}^{(b)} = F_A {\hat {\mbox{\boldmath$n$}}}_{\alpha}.  \label{active_forceB}
\end{eqnarray} 
Eqs. (\ref{active_forceF}) and (\ref{active_forceB}) prescribe a force dipole $F_A\ell_0{\hat {\mbox{\boldmath$n$}}}_{\alpha}$ with $\ell_0{\hat {\mbox{\boldmath$n$}}}_{\alpha}={{\mbox{\boldmath$R$}}}_{\alpha}^{(G)}-{{\mbox{\boldmath$R$}}}_{\alpha}^{(CF)}$ [see also Eq. (\ref{active_stress})].  
Finally, ${{\mbox{\boldmath$F$}}}_{\alpha, {H}}$ and ${{\mbox{\boldmath$N$}}}_{\alpha, {H}}$ are the force and torque exerted on the $\alpha$-th swimmer due to HIs. 
The explicit forms of ${{\mbox{\boldmath$F$}}}_{\alpha, {H}}$, ${{\mbox{\boldmath$N$}}}_{\alpha, {H}}$, and the body force  ${{\mbox{\boldmath$f$}}}_{H}$ can be given in the discretized equations of motion as Eqs. (\ref{Hforce}), (\ref{Htorque}), and (\ref{body_force}), respectively in the next section.

We assume the following form of the interparticle potential: 
\begin{eqnarray}
U^{\mu\nu}(r) = \epsilon(1-\delta_{\mu,f}\delta_{\nu,f}) \biggl(\dfrac{2R}{r}\biggr)^{12},  \label{potentialU}
\end{eqnarray}
where $\epsilon$ is a positive energy constant and $\delta_{\mu,f}$ is the Kronecker delta. 
This form prevents the body part of a swimmer from overlapping on different swimmers but allows overlaps among the flagellum parts. 
The wall-particle interaction potential $W^{\mu}$ is introduced to prevent the penetration of particles through the boundary walls and is assumed to be given as 
\begin{eqnarray}
W^{\mu}(z) = \epsilon \biggl(\dfrac{2R}{z}\biggr)^{12}, \label{potentialW}
\end{eqnarray}
where we assume the same energy constant as that of $U^{\mu\nu}$. 
In Eqs. (\ref{potentialU}) and (\ref{potentialW}), $\mu,\nu=b,f$. 

In our simulations, we make the equations dimensionless by measuring space and time in units of $h$, which is the discretization mesh size used when solving Eqs. (\ref{Navier_Stokes})-(\ref{incompressibility}), and $t_0=\rho h^2/\eta_s$, which is the momentum diffusion time across the unit length. Accordingly, the scaled solvent viscosity is $1$, and the units of velocity, stress, force, and energy are chosen to be $h/t_0$, $\rho h^2/t_0^2$, $\rho h^4/t_0^2$ and $\rho h^5/ t_0^2$, respectively. 
In our simulations, we set $\epsilon=30$ and $F_A=20$. The parameters determining the swimmer's shape are set to be $R=3.2$, $\ell_0=| {{\mbox{\boldmath$R$}}}_\alpha^{(G)}- {{\mbox{\boldmath$R$}}}_\alpha^{(CF)}|=4.5R$ and $\xi=0.5$. In this study, the swimmers' volume fraction is identified as that of the rigid body particles given by $\phi=N{\mathcal V}_\alpha^{(b)}/HL^2$. 

In Ref. \cite{SPM2}, the general scheme deriving the volume-average stress tensor, ${\stackrel{\leftrightarrow}{\mbox{\boldmath$S$}}}$, in the framework of the SPM is provided: ${\stackrel{\leftrightarrow}{\mbox{\boldmath$S$}}}$ is divided into the three parts, ${\stackrel{\leftrightarrow}{\mbox{\boldmath$s$}}}_s$, ${\stackrel{\leftrightarrow}{\mbox{\boldmath$s$}}}_p$, and ${\stackrel{\leftrightarrow}{\mbox{\boldmath$s$}}}_a$, due to the solvent, passive, and active contributions, respectively. 
The passive part is further divided into two parts arising from HIs and the potentials ($U^{\mu\nu}$ and $W^\mu$). 
Such sources to the stress tensor exist without active forces, so we call ${\stackrel{\leftrightarrow}{\mbox{\boldmath$s$}}}_p$ the ``passive'' stress. 
In this Appendix, according to a similar procedure given in Ref. \cite{SPM2}, we derive an expression of the active stress as follows. 
\begin{eqnarray}
{\stackrel{\leftrightarrow}{\mbox{\boldmath$s$}}}_a &=& -\dfrac{1}{L^2 H} \sum_{\alpha=1}^N  F_A{\hat {\mbox{\boldmath$n$}}}_\alpha  \int {\rm d}{{\mbox{\boldmath$r$}}} {{\mbox{\boldmath$r$}}}  \biggl[\dfrac{\Psi^{b}_\alpha ({{\mbox{\boldmath$r$}}})}{{\mathcal V}_\alpha^{(b)}} - \dfrac{\Psi^{f}_\alpha ({{\mbox{\boldmath$r$}}})}{{\mathcal V}_\alpha^{(f)}}\biggr] \nonumber \\
&=& -\dfrac{1}{L^2 H} \sum_{\alpha=1}^N  F_A{\hat {\mbox{\boldmath$n$}}}_\alpha  \biggl({{\mbox{\boldmath$R$}}}_{\alpha}^{(G)} - {{\mbox{\boldmath$R$}}}_{\alpha}^{(CF)}\biggr) \nonumber \\
&=& -\dfrac{1}{L^2 H} \sum_{\alpha=1}^N  F_A\ell_0 
{\hat {\mbox{\boldmath$n$}}}_\alpha {\hat {\mbox{\boldmath$n$}}}_\alpha. \label{active_stress}
\end{eqnarray}
As usual, we may redefine the active stress by making ${\stackrel{\leftrightarrow}{\mbox{\boldmath$s$}}}_a$ traceless by substituting $-({1}/{L^2 H}) \sum_{\alpha=1}^N  F_A\ell_0 {\stackrel{\leftrightarrow}{\mbox{\boldmath$\delta$}}}/3$

In the main text, the viscosity is determined as $\eta=(1/{\dot \gamma}L^2)\int {\rm d}x{\rm d}y\langle \Sigma_{xz}(x,y,\pm H/2)\rangle$, where $\Sigma_{xz}(x,y,\pm H/2)$ is the $xz$ component of the stress tensor at the walls located at $z=\pm H/2$ and $\langle \cdots \rangle$ denotes taking the time average in a steady state. 
Because the relation $\langle S_{xz}\rangle=\langle \Sigma_{xz}\rangle$ holds, $\eta=\langle S_{xz}\rangle/{\dot \gamma}$. As denoted above, ${\stackrel{\leftrightarrow}{\mbox{\boldmath$S$}}}$ are divided into three parts, and then, we have $\eta=\eta_s+\langle s_{p,xz}\rangle/\dot\gamma + \langle s_{a,xz}\rangle/\dot\gamma$. 
The passive part $<s_{p,xz}>/\dot\gamma$ positively contributes to the viscosity, and therefore, the viscosity reduction is entirely due to (weak) alignment of the force dipoles along with the extension direction, $\langle {\hat n}_{\alpha,x}{\hat n}_{\alpha,z}\rangle>0$. 
In the main text, we also discuss how such an alignment of the swimmers in the interior regions reduces the velocity gradient at the boundaries, which is observed as a viscosity reduction.

 We may define the swimmer's Reynolds number as $Re=\rho v_s \ell_0/\eta_s$ with $v_s$ being the average swimming speed. 
In the present simulations, $v_s\sim 0.1$, giving $Re\sim 1$, which is unrealistically large, but may not be problematic for the following reason.  
It is known that unless $Re$ is much greater than unity, the induced flow patterns around a moving particle do not change enough for the inertia effect to significantly change the resultant dynamics and transport properties. 
In Ref. \cite{LB2}, an excellent discussion is presented for colloidal simulations.  
In practice, by using a smaller value of $\rho$, it is possible to reduce $Re$ as much as possible. For this treatment, a smaller time increment is required to ensure the stability of the time integration of the equations of motion, whereas it leads to longer simulation times. In our preliminary simulations at $Re\sim 0.1$, which is still unrealistically large, we confirm that the main results obtained in the present study remain almost unchanged. 

\section{Explicit time-integration algorithm}

Here, following Refs. \cite{SPM,SPM2}, we describe an explicit time-integration scheme for solving model equations as follows. 

The set of physical variables is assumed to be clearly defined at the discrete time step $t_n=n \Delta t$. 

First, we solve Eq. (\ref{Navier_Stokes}) without including ${{\mbox{\boldmath$f$}}}_{H}$ as 
\begin{eqnarray}
{{\mbox{\boldmath$v$}}}^* = {{\mbox{\boldmath$v$}}}_n 
+ \dfrac{1}{\rho}\int _{t_n}^{t_{n+1}} {\rm d}s \biggl(- \rho{{\mbox{\boldmath$v$}}} \cdot {{\mbox{\boldmath$\nabla$}}} {{\mbox{\boldmath$v$}}} +  {{\mbox{\boldmath$\nabla$}}}\cdot {\stackrel{\leftrightarrow}{\mbox{\boldmath$\Sigma$}}_{vis}} + {{\mbox{\boldmath$f$}}}_{A}^{(f)} \biggr)^{\bot},  \nonumber \\
\end{eqnarray}
where ${{\mbox{\boldmath$v$}}}_n$ is the velocity field at $t=t_n$ and $(\cdots)^\bot$ denotes taking the transverse part. 

 Second, we update ${{\mbox{\boldmath$R$}}}_\alpha^{(G)}$ and ${\hat {\mbox{\boldmath$n$}}}_\alpha$ as 
\begin{eqnarray}
{{\mbox{\boldmath$R$}}}_\alpha^{(G)}(t_{n+1}) &=& {{\mbox{\boldmath$R$}}}_\alpha^{(G)}(t_{n}) +  \int _{t_n}^{t_{n+1}} {\rm d}s {{\mbox{\boldmath$V$}}}_\alpha^{(G)} \\
{\hat {\mbox{\boldmath$n$}}}_\alpha(t_{n+1}) &=& {\hat {\mbox{\boldmath$n$}}}_\alpha(t_n) +  \int _{t_n}^{t_{n+1}} {\rm d}s  {{\mbox{\boldmath$\Omega$}}}_\alpha^{(G)} \times  {\hat {\mbox{\boldmath$n$}}}_\alpha. 
\end{eqnarray}
With these updated ${{\mbox{\boldmath$R$}}}_\alpha^{(G)}$ and ${\hat {\mbox{\boldmath$n$}}}_\alpha$,  we also update the variables that determine the swimmer's shapes.  

 Third, the particle velocities and angular velocities are updated by solving Eqs. (\ref{VG}) and (\ref{OG}) as 
 \begin{widetext}
\begin{eqnarray}
{{\mbox{\boldmath$V$}}}_\alpha^{(G)}(t_{n+1}) = {{\mbox{\boldmath$V$}}}_\alpha^{(G)}(t_{n})+\dfrac{1}{M_\alpha}\int _{t_n}^{t_{n+1}} {\rm d}s ({{\mbox{\boldmath$F$}}}_{\alpha, {H}} + {{\mbox{\boldmath$F$}}}_{\alpha, {int}}+ {{\mbox{\boldmath$F$}}}_{\alpha, {A}}^{(b)}+ {{\mbox{\boldmath$F$}}}_{\alpha, {ex}}) , \label{VG2} \\
{{\mbox{\boldmath$\Omega$}}}_\alpha^{(G)}(t_{n+1})={{\mbox{\boldmath$\Omega$}}}_\alpha^{(G)}(t_{n})+{\stackrel{\leftrightarrow}{\mbox{\boldmath$I$}}^{-1}_{\alpha}}\cdot 
 \biggl[ \int _{t_n}^{t_{n+1}} {\rm d}s ({{\mbox{\boldmath$N$}}}_{\alpha, {H}} + {{\mbox{\boldmath$N$}}}_{\alpha, {int}}+{{\mbox{\boldmath$N$}}}_{\alpha, {ex}})\biggr].   \label{OG2}
\end{eqnarray} 
Here, the explicit forms of $\int _{t_n}^{t_{n+1}} {\rm d}s {{\mbox{\boldmath$F$}}}_{\alpha, {H}}$ and $\int _{t_n}^{t_{n+1}} {\rm d}s {{\mbox{\boldmath$N$}}}_{\alpha, {H}}$ are given as 

\begin{eqnarray}
\int _{t_n}^{t_{n+1}} {\rm d}s {{\mbox{\boldmath$F$}}}_{\alpha, {H}} = \int {\rm d}{{\mbox{\boldmath$r$}}} \rho \Psi_{\alpha,n+1}^{(b)} \biggl\{{{\mbox{\boldmath$v$}}}^* -\biggl[{{\mbox{\boldmath$V$}}}_\alpha^{(G)}(t_n)+ {{\mbox{\boldmath$\Omega$}}}_\alpha^{(G)}(t_n)\times \bigl( {{\mbox{\boldmath$r$}}}- {{\mbox{\boldmath$R$}}}_\alpha^{(G)}(t_{n+1})\bigr)  \biggr]  \biggr\} \label{Hforce} 
\end{eqnarray}
and
\begin{eqnarray}
\int _{t_n}^{t_{n+1}} {\rm d}s {{\mbox{\boldmath$N$}}}_{\alpha, {H}} = \int {\rm d}{{\mbox{\boldmath$r$}}} \rho\Psi_{\alpha,n+1}^{(b)}  \bigl( {{\mbox{\boldmath$r$}}}- {{\mbox{\boldmath$R$}}}_\alpha^{(G)}(t_{n+1})\bigr) \times \biggl\{{{\mbox{\boldmath$v$}}}^* -\biggl[{{\mbox{\boldmath$V$}}}_\alpha^{(G)}(t_n)+ {{\mbox{\boldmath$\Omega$}}}_\alpha^{(G)}(t_n)\times  \bigl( {{\mbox{\boldmath$r$}}}- {{\mbox{\boldmath$R$}}}_\alpha^{(G)}(t_{n+1})\bigr) \biggr]  \biggr\},  \label{Htorque}
\end{eqnarray} 
where $\Psi_{\alpha,n+1}^{(b)}$ denotes $\Psi_\alpha^{(b)}({{\mbox{\boldmath$r$}}})$ at $t=t_{n+1}$.  

Finally, we update the velocity field by embedding the rigid body motions in ${{\mbox{\boldmath$v$}}}^*$ through the body force ${{\mbox{\boldmath$f$}}}_{H}$ as 
\begin{eqnarray}
{{\mbox{\boldmath$v$}}}_{n+1} = {{\mbox{\boldmath$v$}}}^* + \dfrac{1}{\rho}\int _{t_n}^{t_{n+1}} {\rm d}s {{\mbox{\boldmath$f$}}}_{H}^\bot. 
\end{eqnarray}
The explicit form of $\int _{t_n}^{t_{n+1}} {\rm d}s {{\mbox{\boldmath$f$}}}_{H}$ is determined to approximately fulfill the rigid body condition inside the swimmers' body region, and it is given by 
\begin{eqnarray}
\int _{t_n}^{t_{n+1}} {\rm d}s {{\mbox{\boldmath$f$}}}_{H} = - \sum_{\alpha=1}^{N}  \rho \Psi_{\alpha,n+1}^{(b)} \biggl\{{{\mbox{\boldmath$v$}}}^* -\biggl[{{\mbox{\boldmath$V$}}}_\alpha^{(G)}(t_{n+1})+ {{\mbox{\boldmath$\Omega$}}}_\alpha^{(G)}(t_{n+1})\times \bigl( {{\mbox{\boldmath$r$}}}- {{\mbox{\boldmath$R$}}}_\alpha^{(G)}(t_{n+1})\bigr)  \biggr]  \biggr\}.  \label{body_force}
\end{eqnarray}
Equations (\ref{Hforce}), (\ref{Htorque}), and (\ref{body_force}) enforce the momentum and angular momentum exchanges between solvent and swimmer's body. The velocity field at the new time step is  
\begin{eqnarray}
{{\mbox{\boldmath$v$}}}_{n+1} = {{\mbox{\boldmath$v$}}}^*\bigl[1-\sum_{\alpha=1}^{N}\Psi_{\alpha,n+1}^{(b)}\bigr] +  \sum_{\alpha=1}^N \Psi_{\alpha,n+1}^{(b)} \biggl[{{\mbox{\boldmath$V$}}}_\alpha^{(G)}(t_{n+1})+ {{\mbox{\boldmath$\Omega$}}}_\alpha^{(G)}(t_{n+1})\times \bigl( {{\mbox{\boldmath$r$}}}- {{\mbox{\boldmath$R$}}}_\alpha^{(G)}(t_{n+1})\bigr)  \biggr] \label{velocity}
   \end{eqnarray}
with
\begin{eqnarray}
\nabla\cdot {{\mbox{\boldmath$v$}}}_{n+1}  &=&0. 
\end{eqnarray}
\end{widetext}
Therefore, within the particle domain ($\Psi_{\alpha,n+1}^{(b)}=1$) the velocity field coincides with the particle velocity as ${{\mbox{\boldmath$v$}}}_{n+1}={{\mbox{\boldmath$V$}}}_\alpha^{(G)}(t_{n+1})+ {{\mbox{\boldmath$\Omega$}}}_\alpha^{(G)}(t_{n+1})\times ( {{\mbox{\boldmath$r$}}}- {{\mbox{\boldmath$R$}}}_\alpha^{(G)}(t_{n+1})) $, while within the fluid domain ($\Psi_{\alpha,n+1}^{(b)}=0$) ${{\mbox{\boldmath$v$}}}_{n+1}={{\mbox{\boldmath$v$}}}^*$. 
In the interface domain ($0<\Psi_{\alpha,n+1}^{(b)}<1$)  the particle velocity smoothly matches the solvent velocity ${{\mbox{\boldmath$v$}}}^*$. That is, the fluid is prevented from penetrating inside the particle domain.  

\section{Evaluation of the hydrodynamic effects on the collision process in our simulations}

As denoted in the main text, when the separation distance between two swimmers is less than $d_0$, these swimmers are assumed to be undergoing a collision; in the present study, we set $d_0=0.7\ell_0$, and the collision time $\Delta t$ is set to be $\Delta t=1.25\ell_0/v_s$, where $\ell_0$ is the swimmer size and $v_s$ is the average swimming speed. This definition of collision is somewhat arbitrary, but we show below that it (with the above-presented sets of $d_0$ and $\Delta t$) can appropriately describe the hydrodynamic collisions. In the following, we make use of the parameter values of  $F_A=20$, $\ell_0=14.4$, $v_s=0.11$, and ${\mathcal V}^{(b)}=355$.

The magnitude of the rotational flow field, $\omega$, induced at a distance $r$ from a  swimmer is approximately $\omega \sim 0.1F_A \ell_0/\eta_s r^3$ \cite{LaugaB}. At the average distance between neighboring swimmers, $r=l_N=({\mathcal V}^{(b)}/\phi)^{1/3}$, $\omega\sim (10^{-1}\phi)\sim 10^{-3}$ for $\phi\sim 0.01$: on average, swimming motions are disturbed by random flows induced by other swimmers at a distance of $\sim l_N$. However, when approaching a specific swimmer, the flow field created by those swimmers deterministically influences their swimming motions. With a similar argument, at $r=d_0$, $\omega\sim 10^{-2}$, and therefore during a collision ($\sim \ell_0/v_s\sim10^2$), the swimmer's trajectory is largely disturbed by HIs as $(\ell_0/v_s)\times \omega \gtrsim 1 $.  In a typical situation with $\dot\gamma=10^{-3}$, for a duration of $\dot\gamma^{-1}$, because $(\pi d_0^2 v_s/{\dot\gamma})\times (1/l_N)^3\sim 10^2 \phi \gtrsim 1$ for $\phi\ge0.01$, at least one ``collision'' may occur. Therefore, in our simulations, the effects of HIs on the swimming motions surpass (or at least compete with) the mean-flow effects for $\dot\gamma\lesssim 10^{-3}$. 

We note that because $(\pi d_0^2 v_s \Delta t)\times (1/l_N)^3\cong 10 \phi \lesssim 1$ in our simulations ($\phi\le0.05$), it is rare that for a duration of $\Delta t$, three or more swimmers are at distances closer than $d_0$. In other words, a hydrodynamic collision can be considered a single event.

\end{document}